\documentclass[runningheads]{llncs}
 \pdfoutput=1 
\usepackage{graphicx}
\usepackage{longtable}
\usepackage{colortbl}
\usepackage{pdfpages}

\usepackage{multirow}
\usepackage{hyperref}
\usepackage{tcolorbox}
\usepackage{orcidlink}
\usepackage{enumitem}
\usepackage{rotating}
\usepackage{environ}
\usepackage{pdflscape}
\usepackage{stmaryrd}
\usepackage{cite}
\usepackage{amsmath,amssymb,amsfonts}
\usepackage{algorithmic}
\usepackage{textcomp}
\usepackage{xcolor}
\usepackage{placeins}
\usepackage{tabularx}
\usepackage{longtable}
\usepackage{fontawesome}
\newcommand\xc{.18}
\newcommand\yc{.82}

\begin{document}
\title{Setting AI in context\thanks{This project has received funding from the European Union’s Horizon 2020 research and innovation program under grant agreement No 957197.}:\\ A case study on defining the context and operational design domain for automated driving
}
\titlerunning{Setting AI in context}
%
\author{Hans-Martin Heyn\inst{1,2}\orcidID{0000-0002-2427-6875},\textsuperscript{\faEnvelopeO} \and
Padmini Subbiah\inst{1} \and
Jennifer Linder\inst{1} \and
Eric Knauss\inst{1,2}\orcidID{0000-0002-6631-872X} \and
Olof Eriksson\inst{3}}
\authorrunning{H.-M. Heyn et al.}
%
\institute{Chalmers University of Technology, SE-412 96 Gothenburg, Sweden \and
University of Gothenburg, SE-405 30 Gothenburg, Sweden \and
Veoneer Sweden AB, SE-103 02 Stockholm, Sweden}
%
\maketitle              

\begin{abstract}
\noindent \textbf{[Context and motivation]} 
For automated driving systems, the operational context needs to be known in order to state guarantees on performance and safety. 
The operational design domain (ODD) is an abstraction of the operational context, and its definition is an integral part of the system development process. 
\noindent \textbf{[Question / problem]} There are still major uncertainties in how to clearly define and document the operational context in a diverse and distributed development environment such as the automotive industry. 
This case study investigates the challenges with context definitions for the development of perception functions that use machine learning for automated driving. 
\noindent \textbf{[Principal ideas/results]} Based on qualitative analysis of data from semi-structured interviews, the case study shows that there is a lack of standardisation for context definitions across the industry, ambiguities in the processes that lead to deriving the ODD, missing documentation of assumptions about the operational context, and a lack of involvement of function developers in the context definition.
\noindent \textbf{[Contribution]} The results outline challenges experienced by an automotive supplier company when defining the operational context for systems using machine learning. 
Furthermore, the study collected ideas for potential solutions from the perspective of practitioners.

\end{abstract}

%
%
%
\section{Introduction}
Automated driving systems (ADS) rely on machine learning (ML) especially for cognition tasks and sensor fusion. 
Machine learning, as part of artificial intelligence (AI), experiences an advent of methods, tools, and applications, especially due to breakthroughs in applying deep neural networks to machine learning problems. \par
The growing interest in the development of systems that can take control of driving is accompanied by concerns regarding safety, i.e., assuring that the ADS is able to operate safely and as expected in the desired operational context~\cite{Reschka2012,Shalev-Shwartz2017}. \par
At present, an answer to the safety concern is to keep the context for automated driving very limited, for example to factory sites, harbours, or mining operations. 
By developing ADS for many different, limited, contexts, the hope is to take the experiences and lessons learned from these limited contexts and apply them to a wider context, such as automated driving on highways.
The challenge is, that a wider context will cause a formidable grow in possible scenarios and situations. 
With the current processes and methods that were developed (or naturally grown) for limited contexts, it will be difficult to capture and describe all the possible scenarios that the vehicle can encounter in a wider context.
Another challenge is owed to the way of working in the automotive industry. 
Much of the product development is done either solely by a supplier company, or in cooperation with the original equipment manufacturer (OEM).
This requires efficient and correct processes for communication of information regarding the system context between the customers, the OEM, and the supplier companies. \par
This case study investigates qualitatively the current challenges and solution ideas of a Tier 1 supplier\footnote{a Tier 1 supplier develops and sells products and solutions directly to an OEM} regarding the definition of context and operational design domain from use cases for systems of automated driving that incorporate machine learning. Interviews with employees in a variety of different positions at the supplier company are the main source of data for this study in addition to data collected at different OEMs and partner companies. The findings were triangulated with background literature and a focus group validated the findings from the interview study. The study finds deficits in the standardisation of context definitions and ODDs, uncertainty and lack of transparency in processes for context definitions, insufficient documentation of context assumptions, and too little participation of function developers in use case interpretation and context definitions. \par
Section~2 of this paper describes the background and briefly the history of context definitions for computer systems and provides a problem definition and research questions. Section~3 explains the applied methodology. Section~4 presents the validated findings of the study. Section~5 includes the triangulation of findings to the background literature, a summary, and a discussion of the main findings.
\section{Background}
A system's desired behaviour and responsibilities are often described through textual use cases. 
They state how a system reacts to different situations with as little text as possible, but also clearly convey the reactions to these situations \cite{Cockburn2000}. 
The task of a requirement engineer is to translate the use cases into requirements for the system. 
Different requirements concern different parts of the system: Examples of these are functional requirements, quality requirements, safety requirements, etc. 
However, some of the requirements are linked to a specific context in which they are valid in \cite{Knauss2014}. 
At design time, contextual attributes that can change at run time need to be identified in order to avoid uncertain or even undesired behaviour of the system at run time. 
A contextual requirement then forms a tuple of desired behaviour (requirement) and the required state of the contextual attributes \cite{Knauss2016}. 
However, considering every contextual attributes and their changes at design time requires a complete understanding of the operational environment which is not feasible for complex or even chaotic behaving environments \cite{Ramirez2012}. 
What does it mean to talk about the "context of a system"? The Oxford Learner's Dictionary defines "context" as:

\begin{tcolorbox}
\footnotesize
"[Context is] the situation in which something happens and that helps you to understand it" \cite{CambridgeContext2021}.
\end{tcolorbox}

\noindent For computer systems, Dey provided a more specific definition of context:

\begin{tcolorbox}
\footnotesize
"Context is any information that can be used to characterise the situation of an entity.
An entity is a person, place, or object that is considered relevant to the interaction between a user and an application, including the user and applications themselves" \cite[p.~4]{Dey2001}.
\end{tcolorbox}

\noindent Researchers in systems engineering extended the definition of context of a system by including the environment in which the system shall operate \cite{Brown1996,Henricksen2004}. 
Chen et al. further extended the definition of context by adding system capabilities and the situational roles, beliefs, and intentions of people engaged with the system to the definition of context:

\begin{tcolorbox}
\footnotesize
"[Context is] information about a location, its environmental attributes (e.g., noise level, light intensity, temperature, and motion) and the people, devices, objects and software agents it contains. 
Context may also include system capabilities, services offered and sought, the activities and tasks in which people and computing entities are engaged, and their situational roles, beliefs, and intentions" \cite[p.~1-2]{Chen2003}.
\end{tcolorbox}

\noindent Nemoto et al. introduces "spatial-temporal elements", and thus adds a temporal dimension to the context \cite{Nemoto2015}. 
The development in vehicle automation increased the discussion around context definition for computer systems. 
Because of the temporal dimension, the context is highly dynamic, and it is important to find a systematic way to describe and confine the context of a vehicle automation system \cite{Soultana2019}. 
Traditionally, scenarios are created with the aim to represent typical driving situations in a given context. 
Damak argues that it is difficult to capture all relevant contextual elements in a scenario-based approach. 
He therefore proposed to build the scenarios in discrete stages based on different context elements, such as use case, environment, road infrastructure, and traffic objects \cite{Damak2020}. \par 

Besides context definition, the term "operational design domain" (ODD) has become popular when discussing capabilities and limitations of vehicle automation systems. 
The SAE standard J3016 \cite{SAE2018} introduced a widely adopted classification of driving automation into six levels: Level 0 indicates no automation at all, and thus an ODD is not applicable. 
Level 1 to 4 indicate different levels of automation, from merely driver assistance (Level 1) to full automation in a predefined environment (Level 4). 
On the highest level of automation (Level 5), the ODD is "unlimited". The standard defines the ODD as:

\begin{tcolorbox}
\footnotesize
 "[The ODD describes the] operating conditions under which a given system for driving automation or feature thereof is specifically designed to function, including, but not limited to, environmental, geographical, and time-of-day restrictions, and/or the requisite presence or absence of certain traffic or roadway characteristics" \cite[p.~14]{SAE2018}.
\end{tcolorbox}

\noindent This definition leaves much room for interpretation of what constitutes an ODD and how to argue for completeness. 
Many proposals for what constitutes an ODD have been put forward in recent years (for example a discrete list of ODD items \cite{NHTSA2017}, a detailed ontology of road structures \cite{Czarnecki2018}, runtime monitoring requirements \cite{Colwell2018}, another categorised list of ODD items \cite{Koopman2019}, or internal system capabilities \cite{Gyllenhammar2020}). 
Still, there is a lack of a common definition for the ODD, which creates challenges in communication and collaboration between the stakeholders of the system \cite{Thorn2018}. \par
\vspace{2mm}
\noindent\textbf{Problem Definition:}
A use case assumes a context, and the resulting requirements will only be valid in that assumed context. 
For safety relevant systems, the dependencies of the requirements on the context is specifically obvious: A specific level of safety is only guaranteed in a clearly specified and tested ODD. 
Outside of the ODD, the behaviour of the system cannot be guaranteed to any safety level. 
The problem is, that there is neither a clear understanding of how to define the context in which a system shall operate, nor is there a common definition for an ODD. 
Ideally, the use cases for a system should include information about the context in which it shall operate within. 
However, use cases are often quite broad and non-specific, which requires the practitioners to interpret how the context for the requirements needs to be defined from the use case. 
Especially for adaptive systems, it is important to relate requirements to a specific context, because the context might change while the system is active \cite{Knauss2016}.\par
\vspace{2mm}
\subsection{Research Design}
The problem of unclear context definitions is especially problematic for the development of functions for automated driving that use some form of AI: Without a clearly outlined context of the desired use case, it will be impossible to refine a testable operational design domain in which performance and safety aspects can be guaranteed or to find the right data sets for training and validation of the AI.
This study investigates and explores the current status, challenges, and possible improvements for deriving context definitions from use cases for automated driving and advanced driver assistance systems. 
The study is carried out as a case study at a Tier 1 automotive supplier that develops and provides sensor systems for automated driving systems. \par
The aim of this study is to provide views and information on challenges with deriving context definitions and ODDs from use cases, in the setting of a Tier~1 supplier providing machine learning supported sensor systems for automated driving. 
This empirical study does not provide a set of solutions for the challenges. \par
\vspace{2mm}
\noindent\textbf{Research Questions:}
Following the research approach for empirical case studies outlined in \cite{Creswell2017}, the research questions that guide this study are formulated as open-ended questions. 
They focus on the previously described central phenomenon of deriving context definitions and ODDs from use cases. \par
\vspace{2mm}
\noindent \textbf{Research Question 1 (RQ1):} What is the current understanding of context definitions? \par
\noindent \textbf{Research Question 2 (RQ2):} What are the challenges with deriving context definitions from use cases? \par
\noindent \textbf{Research Question 3 (RQ3):} Which support would be appropriate for deriving context definitions from use cases?

\section{Methodology}
\begin{figure}[b]
    \centering
    \includegraphics[width=\linewidth]{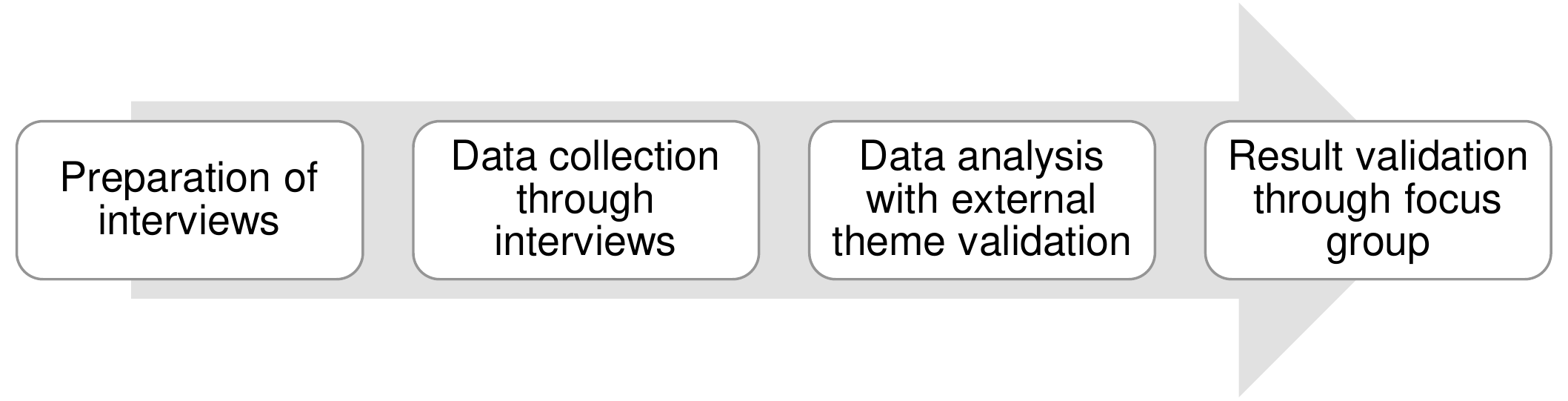}
    \caption{Overview of the applied methodology}
    \label{fig:method}
\end{figure}

Figure~\ref{fig:method} gives an overview of the applied methodology, which consists of four steps: Preparation of interviews, data collection through interviews, data analysis, and result validation.

\subsection{Preparation of interviews}
The aim of the data collection was to illuminate the situation and challenges with context definitions as they are experienced primarily from the perspective of a Tier 1 supplier. 
The reason behind choosing a Tier 1 supplier as site of the investigation is that automated driving functions are part of highly complex systems that are primarily developed in cooperation between an OEM and its Tier 1 suppliers \cite{Pfeffer2019}. Because of the cooperation with OEMs, public authorities and other research organisations in the development of automated driving functions, we chose to collect some of the data from OEMs, a public traffic regulation authority and a research company. All of these parties have worked in collaboration with the Tier 1 supplier in the past on automatic driving projects.\par
\vspace{2mm}
\noindent \textbf{Sampling strategy:}
This empirical study follows a maximum variation strategy for sampling \cite{Creswell2017}.
The participants were chosen purposefully to represent a wide variety of experiences and positions involved in the development of automated driving functions that use machine learning \cite{Palinkas2015}.
To simplify the filtering of suitable candidates, we defined four position groups:\par

\vspace{2mm}
\noindent \textbf{Positions with a high level perspective:} system managers, system engineers, system architects, and system designers; \par
\noindent \textbf{Positions dealing primarily with requirements:} requirements engineers and (public) policy makers; \par
\noindent \textbf{Positions with a customers / end-user focus:} product owners and function owners; \par
\noindent \textbf{Positions with a clear focus on development:} function developers and system developers which develop the function/system based on given specifications.\par
\vspace{2mm}

\noindent The aim of the sampling strategy was to have representation of each of these groups to ensure a view on the entire system development chain. 
Seven participants from two Tier 1 supplier companies located in the United States and Sweden were interviewed. 
To add the OEM's perspective to the sampling data, we also interviewed four participants from three different OEMs (one OEM each from Sweden, China, and Japan). 
In order to increase diversity among the participants and to reduce company induced bias in the results, one additional person could be interviewed from a Japanese automotive technology research company, and one person participated from the Swedish Transport Administration (Trafikverket). 
Altogether 13 interviewees participated in this study. 
A full list of participants and their respective roles is given in Table \ref{tab:Participants}\footnote{Note that due to privacy concern, we intentionally chose not to reveal the respective company}.

\begin{table*}[t]
\centering
\caption{Participants of the case study}
\label{tab:Participants}
\begin{tabular}{cll}
\hline
\rowcolor[HTML]{C0C0C0}
\multicolumn{1}{c}{\cellcolor[HTML]{C0C0C0}\textbf{\begin{tabular}[c]{@{}c@{}}Inter-\\ viewee\end{tabular}}} & \multicolumn{1}{c}{\cellcolor[HTML]{C0C0C0}\textbf{Role}} & \multicolumn{1}{c}{\cellcolor[HTML]{C0C0C0}\textbf{Experience}} \\ \hline
A & Group Manager &  ADAS features and collision avoidance features. \\
\rowcolor[HTML]{EFEFEF} 
B & Functional Developer &  ADAS features. \\
C & System Engineer &  Planning and control for safety critical issues.\\ 
\rowcolor[HTML]{EFEFEF} 
D & Researcher & Innovation for sensors and systems. \\
E & Developer & Algorithms for obstacle detection. \\
\rowcolor[HTML]{EFEFEF} 
F & \begin{tabular}[l]{@{}l@{}}Functional Developer and\\ Functional Safety Engineer\end{tabular} & ADAS feature development. \\
G & Project Manager & ADAS vehicles.  \\
\rowcolor[HTML]{EFEFEF} 
H & Researcher & Data management and computer vision. \\
I & Product Owner & Ground truth systems. \\
\rowcolor[HTML]{EFEFEF} 
J & Technical Lead & AI and machine learning projects. \\
K & \begin{tabular}[l]{@{}l@{}}Technical Specialist \\ Adaptive Cruise Control \end{tabular} & \begin{tabular}[l]{@{}l@{}} Blind spot detection, lane changing,\\ adaptive cruise control, collision avoidance.\end{tabular} \\
\rowcolor[HTML]{EFEFEF} 
L & Functional Safety Manager & Functional safety work and documentation. \\
M & Researcher & Standardisation of safety methodologies. \\

\hline
\end{tabular}
\end{table*}

\subsection{Data collection through interviews}
The interview questions, collected in an interview guide only available to the interviewers, were formulated based on the a-priori formulated research questions.  
It was divided into three sections: The first section aimed at identifying the participant's current role and experience. 
The second section established some ground concepts with the interviewee. 
This was done to avoid misunderstandings, for example due to different definition of terms. For example, the interviewees were given the SAE's description of the operational design domain and providing examples for different context definitions. 
For each of these examples, the interviewees were asked to provide their opinion on applicability and problems with the provided examples. 
The third section explored the process of deriving context definitions from use cases. 
The aim was to investigate the multiple facets of the process, including identifying the main concerns, describing of what works well in the process, and registering possible improvements. 
In some interviews some additional follow-up questions were included.
The interviews were conducted individually with each participant for about one hour remotely via Microsoft Teams or Zoom. 
One interviewer asked the questions, while the second interviewer observed and took notes. 
Each interview started by presenting information about the study's objective.

\subsection{Data analysis}
\begin{figure}[b]
    \centering
    \includegraphics[width=0.75\linewidth]{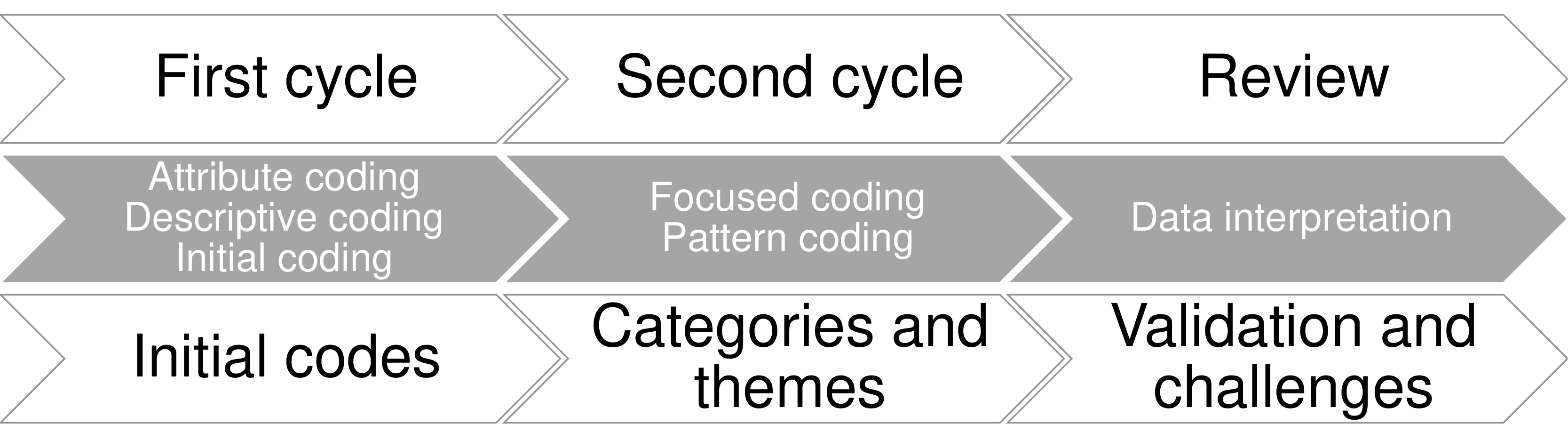}
    \caption{Steps of the data analysis}
    \label{fig:analysis_steps}
\end{figure}

Except for one interview, all the interviews were recorded and transcribed. 
For one interview, both the interviewer and an observer took notes. 
The data analysis consisted of the three steps illustrated in Figure~\ref{fig:analysis_steps}. 
The coding strategy was determined through pilot coding, conducted by two researchers independently and the results evaluated and discussed with all authors.\par

\vspace{2mm}
\noindent \textbf{First cycle coding:}
The first cycle coding consisted of three steps: With attribute coding, meta information such as the role of the interviewees and work experiences were recorded. 
Descriptive coding allowed for developing codes that represent different topics of the statements given by the interviewees. 
Finally, initial coding, as suggested by Salda{\~{n}}a, was used to highlight and understand the interviewees' thoughts \cite{Saldana2013}. \par
\vspace{2mm}
\noindent \textbf{Second cycle coding:}
With focused coding, the initial codes from the first cycle were divided into broad categories. 
These broad categories were further split into subcategories. 
Afterwards, pattern coding was used to find emerging themes among the subcategories. 
Finally, each statement from the interviews were assigned to one of the created subcategories. 
A second independent group of three researchers validated the found themes and the assignment of both challenges and potential solutions mentioned by the interviewees to the themes.
With this step, it was tried to reduce bias in the selection of the themes. \par
\vspace{2mm}
\noindent \textbf{Validation of findings:}
This study uses the "member checking" validation strategy as described in \cite[ch.~9]{Creswell2014} to assess the validity and accuracy of the results. 
For this purpose, a focus group with four interviewees from the data collection phase was conducted. 
The session was conducted remotely using Mentimeter, a web-tool that allows for interactive questioning of the participants. 
The questions for the focus group were prepared a-priori.
The themes related to the challenges were presented to the focus group, and the participants were asked to either agree or disagree with the challenges. 
Furthermore, the participants were asked to discuss the themes related to the proposed solutions of the challenges, and they were asked to rank the themes according to the participant's opinion on how important the solution of a particular challenge is.
\section{Results}
This section presents the results based on the data obtained through the interview sessions. 
The section is divided into subsections presenting each research questions individually because the research questions built upon each other. 

\subsection{RQ1: What is the current understanding of context definitions?}
The first 15 minutes of the interviews were used to establish an overview of the interviewee's understanding of the terms context and ODD. \par
\vspace{2mm}
\noindent \textbf{What is context?} The interviewees were asked to elaborate on their understanding of what is meant by context definition in relation to automated driving. 
All interviewees were not entirely sure what context definition describes in relation to automated driving systems.
Two interviewees considered the context definition and ODD to be identical in that they both describe environmental conditions in which the system is designed to operate:

\begin{tcolorbox}
\footnotesize
“[...] it’s kind of the same things as the ODD is describing but context contains all the possible combinations, if you like, of where this is going to operate.”
\end{tcolorbox}

\noindent However, the majority of interviewees did not see context and ODD as identical definitions. 
Instead, they stated that the ODD is a form of representation of the context. \par

\vspace{2mm}
\noindent \textbf{What entails the context?} All interviewees describe the context as a dynamic, and rather wide entity, that should include situations, scenarios, and the environments in which a system operates.

\begin{tcolorbox}
\footnotesize
"But it is a way to define a situation, or define a system, or defining [...] a scenario, or an environment [...]. So that is what I would call a context."
\end{tcolorbox}

\noindent One interviewee includes also the functional state of the vehicle in the context.
All interviewees stated that knowledge and a clear description about the context is important for validation, safety, and security of the system. 
If assumptions about the context of a system are made, they need to be clearly communicated as assumptions, which is according to the interviewees not always the case. \par
\vspace{2mm}
\noindent \textbf{What is the ODD?} All interviewees agreed with the SAE J3016 definition of the ODD \cite[Page 14]{SAE2018}. 
Although SAE J3016 provides a definition of the ODD, all interviewees mention that there is lack of standardisation of the format of an ODD. 
A majority of interviewees added that a description of a design domain beyond the actual ODD, in which the system's performance is degraded but still safe, is necessary in addition to the ODD. As an example, interviewee F described the process in place for deriving the ODD: After the use case is defined, an exploratory search starts to identify the context in which "the use case is actually happening". Once they established an understanding of the context, the internal conditions of the vehicle and the external environment are analysed for the given context. The information about internal conditions of the vehicle and external environment states are what defines the ODD. In this process, the ODD is derived from the use cases via an exploratory search of the context. Interviewee K however describes the process for defining the ODD different: After an initial ODD is defined, the ODD is analysed and use cases and requirements are derived based on the initial ODD. An iterative process is started to adjust ODD and use cases "back and forth" until ODD and use cases comply with each other. This is done by first reducing the number of use cases to the most relevant ones that fit into the initial ODD. Then, they try to get a better understanding of the capabilities of the system, which allows them to widen the initial ODD and to take up more use cases gradually. In contrast to the process described by interviewee F, here they start with the ODD and derive the use cases based on the targeted ODD.\par

\begin{tcolorbox}
\footnotesize
"[...] in the ODD you have to describe it as sort of a graceful degradation of the system when you go outside it's never included".
\end{tcolorbox}
\vspace{2mm}
\noindent \textbf{What entails the ODD?} A majority of interviewees claimed that the SAE definition is incomplete. 
A major missing aspect that was mentioned is the internal state of the vehicle, i.e., capability of sensors and actors.
Furthermore, the drivers' behaviour in and around the vehicle should be described in the ODD (e.g. can the system operate with an intoxicated or distracted driver?). 
Ten interviewees reminded that the road and lane conditions should explicitly be highlighted in the ODD, as they play a major role in the correct function of automated driving systems. \par
\vspace{2mm}
\noindent \textbf{What is the difference between context and ODD?} All interviewees associate the ODD with safety and performance guarantees. The context is associated with validity of requirements derived from a given use case. An interpretation of the interviewees' answers can be, that the ODD is an abstraction, or mode, of the context. Similar to the World-Machine Model described by Jackson in \cite{Jackson1995}, the ODD can be interpreted as an abstraction of the context in which a given system can testable operate with desired characteristics, such as safety, reliability, and performance. And in some cases, as described by two of the interviewees, the ODD as abstraction and the context itself, seem identical for a given use case. This can occur, for example, if the the context derived from the use case is limited enough such that the ODD can completely embrace the operational context.


\subsection{RQ2: What are the challenges with deriving context definitions from use cases?} \label{subsec:RQ2}
This section describes the major challenges through themes identified from the interviews. Three areas of challenges illustrated were identified from the interview sessions and validated by the focus group. They are "deriving context definitions", "process and communication of context definition", and "deriving the ODD from context definition". For each area of challenge, only the themes validated with a simple majority by the focus group members are presented in Table~\ref{tab:Challenges}. A full list of themes can be made available upon request.

\begin{center}
\scriptsize
\begin{longtable}{{m{\xc\textwidth}m{\yc\textwidth}}}
\caption{Themes relates to challenges with deriving context definitions.}
\label{tab:Challenges} \\
\hline
\multicolumn{1}{c}{\cellcolor[HTML]{C0C0C0}\textbf{Theme}} & \multicolumn{1}{c}{\cellcolor[HTML]{C0C0C0}\textbf{Description}} \\ \hline
\endfirsthead

\hline
\multicolumn{1}{c}{\cellcolor[HTML]{C0C0C0}\textbf{Theme}} & \multicolumn{1}{c}{\cellcolor[HTML]{C0C0C0}\textbf{Description}} \\ \hline
\endhead

\hline \multicolumn{2}{r}{{Continued on next page}} \\ \hline
\endfoot

\hline
\endlastfoot

\multicolumn{2}{l}{\cellcolor[HTML]{656565}\color[HTML]{FFFFFF}Deriving context definitions} \\
\textbf{Difficult to \newline describe \newline context} & There is a lack of terminology for describing non-numeric parameters within the context, like the weather: It is for example not clear what "in good weather conditions" actually means. Therefore, non-numeric parameters are difficult to compare between different context descriptions. Furthermore, the environment is often dynamic and containing unknown unknowns, which can change the context of a system unpredictably. \\
\rowcolor[HTML]{EFEFEF} 
\textbf{Lack of \newline standard} & A common language for context definitions is lacking, which makes it difficult to work on a system or product in different countries, companies, or even teams. Unlike requirement specification, there is no correspondent context specification, which results in ambiguities in the context in which requirements are valid.\\
\textbf{Lack of \newline transparency} & A lack of transparency in use case creation and requirement negotiation leads to challenges when defining the desired and feasible context of the system. It is seen as important that the function developers obtain more knowledge about the use case in order to understand the necessary context in which a function / system shall operate. Furthermore, it was stated that there is no good practise in determining if a system still operates within its designated context. \\
\hline
\multicolumn{2}{l}{\cellcolor[HTML]{656565}\color[HTML]{FFFFFF}Defining ODDs} \\
\rowcolor[HTML]{EFEFEF} 
\textbf{Lack of \newline arguments for \newline completeness} & It is difficult to know when an ODD captures all relevant scenarios and elements, mainly because there is no standardised method or template for determining if the ODD is complete. \\
\textbf{Difficult to \newline capture all \newline scenarios in \newline ODD} & To enumerate all possible scenarios in the ODD is impossible. It is difficult to determine, which scenarios the ODD should entail, and which scenarios are not required to be captured by the ODD. Especially "edge case" scenarios are difficult to describe in the ODD, because many assumptions are necessary in these scenarios, which are often not well documented in the context description. \\
\rowcolor[HTML]{EFEFEF} 
\textbf{Hard to \newline understand \newline context \newline definitions} & Function developers are not always involved in defining the context definitions, which makes it difficult for them to develop an ODD that fits the desired context.   \\
\textbf{Lack of \newline standard \newline for ODD} & ODD tends to mean different things, which makes it difficult to understand what an ODD shall entail. Different OEMs, and even different teams within an OEM, have different approaches to define the ODD, which creates confusion for Tier 1 suppliers. The lack of a standardisation has been mentioned by all interviewees as a major obstacle.\\
\rowcolor[HTML]{EFEFEF} 
\textbf{Overly \newline cautious} & A consequence of not knowing the right context of the system is that the ODD will be overly cautious. Developers will start with a too strict and too limited ODD, and only expand it once safety has been proven within small extensions. This can lead to unnecessary long developing and testing times, or overly cautious systems.\\
\hline
\multicolumn{2}{l}{\cellcolor[HTML]{656565}\color[HTML]{FFFFFF}Process and communication} \\
\textbf{Assumption \newline not \newline documented} & A concern mentioned was that assumptions about the context and in the requirements are not being properly documented as such. In many cases it is necessary to make assumptions, but they must be clearly documented as such. A typical context assumption for a function is to assume a linear behaviour of some measured dynamic. \\
\rowcolor[HTML]{EFEFEF} 
\textbf{Insufficient \newline involvement \newline of function \newline developers} & Function developers are detached from the overall picture, because they are not involved enough in defining requirements and context definitions. This makes it hard for them to understand the context in which the system is supposed to function. \\
\textbf{Lack of \newline feedback} & Sometimes changes in the context, and even requirements, of a system are only discovered during the development. A feedback loop is often missing to verify if these changes in context are acceptable. \\
\rowcolor[HTML]{EFEFEF} 
\textbf{Misinter- \newline pretation of \newline requirements} & Textual requirements and context definitions can be misinterpreted by different peoples with different views on the system. Often, the person writing the requirements and context definitions has no direct contact to the person implementing them. \\
\textbf{Too difficult \newline process} & The process of deriving contextual information and requirements from use cases was described as being "blurry and unsharp by nature". It was also mentioned, that a common structured process is either missing or too complex for deriving both requirements and context definitions. \\
\hline
\end{longtable}
\end{center}
\FloatBarrier

\subsection{RQ3: Which support would be appropriate for deriving context definitions from use cases?}
For each of the three areas of challenges described in Section \ref{subsec:RQ2} the interviewees were asked to suggest improvements. Out of the interviews, themes were identified and presented to the focus group for validation. All themes that achieved a simple majority vote in the focus group are presented in Table \ref{tab:Improvements}. 

\begin{center}
\scriptsize
\begin{longtable}{{m{\xc\textwidth}m{\yc\textwidth}}}
\caption{Themes related to improvement ideas for deriving context definitions.}
\label{tab:Improvements} \\
\hline
\multicolumn{1}{c}{\cellcolor[HTML]{C0C0C0}\textbf{Theme}} & \multicolumn{1}{c}{\cellcolor[HTML]{C0C0C0}\textbf{Description}} \\ \hline
\endfirsthead

\hline
\multicolumn{1}{c}{\cellcolor[HTML]{C0C0C0}\textbf{Theme}} & \multicolumn{1}{c}{\cellcolor[HTML]{C0C0C0}\textbf{Description}} \\ \hline
\endhead

\hline \multicolumn{2}{r}{{Continued on next page}} \\ \hline
\endfoot

\hline 
\endlastfoot

\multicolumn{2}{l}{\cellcolor[HTML]{656565}\color[HTML]{FFFFFF}Improvement ideas for deriving context definitions} \\
\textbf{More \newline diverse data} & According to one interviewee, a more diverse set of sensor data of the environment allows for easier interpretation and limitation of the context in which the system operates.\\
\rowcolor[HTML]{EFEFEF} 
\textbf{Standardised \newline approach} & All interviewees suggested to standardised context definition to ease cooperation between different teams and companies. \\
\hline
\multicolumn{2}{l}{\cellcolor[HTML]{656565}\color[HTML]{FFFFFF}Improvement ideas for defining the ODD} \\
\textbf{Automatic \newline tool for \newline deriving ODD} & The described improvement would encompass a tool that can take as input the context, requirements based on the context and test cases. It would then propose an appropriate ODD, that is valid in the desired context, entails all requirements, and is verified through test cases. \\
\rowcolor[HTML]{EFEFEF}
\textbf{Complete- \newline ness criteria \newline for ODD} & One interviewee suggested that an explicit method and criteria are needed to proof that the ODD is complete and correct. \\
\textbf{Information \newline should be \newline described \newline clearly} &  The ODD should contain more information about the context of the system. According to one interviewee, this would improve identifying wrong assumptions about the context early. \\
\rowcolor[HTML]{EFEFEF}
\textbf{Measure of \newline exposure to \newline a safety event} & One interviewee explained that the hazard and risk assessment to evaluate the dimensions of the ODD from a safety viewpoint. The HARA includes assumptions about the context, and especially the exposure to a hazardous context can then be used to decide on the ODD. \\
\textbf{Standardised \newline process} & All interviewees suggested to improve the standardisation of processes for defining the ODD. \\
\hline
\multicolumn{2}{l}{\cellcolor[HTML]{656565}\color[HTML]{FFFFFF}Improvement ideas for the process and communication} \\
\rowcolor[HTML]{EFEFEF}
\textbf{Better \newline continuous \newline improvement} & Automated driving is a new technology, which needs to evolve continuously. The processes for deriving relevant artefacts, such as the context definition and ODD need to evolve together with the technology. \\
\textbf{Faster \newline feedback} & One suggested improvement was to derive requirements faster, and creating faster feedback loops between the stakeholders. This expedites also the definition of the context, because assumptions can be made and verified faster. \\
\rowcolor[HTML]{EFEFEF}
\textbf{Improved \newline leveraging from \newline other contexts} & Typically, automated driving systems are developed for and tested in confined areas, such as factory areas or harbours. Lessons are learnt in these confined contexts, and it is important to be able to leverage the knowledge from these confined contexts into new contexts. \\
\textbf{Improved \newline integration \newline into \newline SAFe setup} & It is perceived that requirement engineering is not well integrated in scaled agile frameworks (SAFe), which hinders an efficient development and handling of requirements, and consequently context definitions.\\
\rowcolor[HTML]{EFEFEF}
\textbf{Involvement \newline of function \newline developers} & Function developers should be more involved in the requirement engineering process, including context definition and deriving the ODD. On one hand, the function developers would get a better understanding of the requirements and context. On the other hand, they can contribute with deep knowledge about the used technologies, which eases the understanding of the the technology's capability for desired contexts. \\
\hline
\end{longtable}
\end{center}
\FloatBarrier
\section{Discussion}
\subsection{Triangulation with background literature}
Already in 2001, Dey described that for computing environments there is only "an impoverished understanding of what context is and how it can be used" \cite{Dey2001}. 
The results from the interviews show that although the understanding about context definitions have increased over 20 years, it is still difficult to use context definitions in practise. 
A main reason for this seems to be lack of standardisation and processes when dealing with context definitions, and ODDs as abstraction thereof. Some attempts for standardisation of an ODD taxonomy have been attempted recently, such as described in  \cite{Irvine2021,Schwalb2021} or through pending standards such as ISO/TR 4804:2020 \cite{ISO4804}.
This lack of standardisation in regards to ODDs (as abstraction of context definitions) has also been described by Gyllenhammar et al. \cite{Gyllenhammar2020}. 
The difficulty to capture all scenarios in an ODD has also been described in \cite{Koopman2019} and arguing for completeness has been discussed in \cite{Henricksen2004}.
A theme mentioned by all interviewees was that the process for deriving context definitions is difficult in the sense that it contains too many uncertainties. 
Damak et al. identified this difficulty as well, and developed a method to adopt architectural decisions for automated driving systems to the operational context \cite{Damak2020}. 
Thron et al. observed challenges in the communication of ODDs between stakeholders of a system \cite{Thorn2018}. 
A majority of interviewees, including all function developers, described a lack of transparency in requirement negotiations (and context definition) for desired use cases, which indicates that the communication problem is not solved.

\subsection{Discussion and main findings}
Keeping the background literature in mind, we argue for four main findings that can trigger future investigations:
Firstly, we identified confusion in the definitions of context, and operational design domain. 
The connection between the context definition and operational design domain is ambivalent and requires more clarification, for example through standardisation. 
Furthermore, we identified a lack of clear processes leading to context definitions and ODDs. 
Although many attempts of creating some form of standard or template process exist, especially for the case of ODD, there is no clear picture in our case. Major problems are arguing for completeness and lack of stakeholder involvement.
Also, we noticed problems when defining and documenting assumption about the context. Interviewees reported that assumptions about the context are not documented as such, and therefore it is difficult to differentiate assumptions about the context from facts during function development.
Lastly, we observed a disconnection of the function developers from the requirement engineering, which also includes the context definition from use cases. 
Especially with the introduction of more agile frameworks, it is beneficial to move parts of the requirement engineering towards the function developers, including defining context from use cases and deriving of ODDs.
Applying machine learning in systems for automated driving systems requires that the context of the systems can be clearly defined and described in well working processes. 
Machine learning is a key technology for perception systems in automated driving. 
Often implicitly, by selecting data sets for training and validation, machine learning models are limited to the context represented in these data sets. 
Specifically, a machine learning system requires not only an understanding of the desired behaviour (given through use cases, and functional requirements), but also of the context in which the system operates. The context is important, because it defines both the necessary training and necessary testing scenarios of the system. The training scenarios, and in most learning scenarios the training dataset, define the final behaviour of the machine learning system. Desired behaviour and context are therefore closely intertwined, and that might be a reason for the difficulties observed in defining the context and consequently the ODD for a machine learning system. Should the desired behaviour in form of use cases be defined first, or should the desired context be first? We saw in the answers that there is no clear picture on the order and the processes of defining use cases, context and ODD. This interlacement of use case and context, and the lack of established processes, could be the reason for the overly cautious definitions of ODDs that was reported in the interviews.
We argue that a better understanding is needed how the context influences the desired behaviour of a system with machine learning components. This relation between desired behaviour and context needs to be made explicit, in order to understand the consequences on the desired behaviour of context changes. Based on the explicit definition of the context and its relation to the desired behaviour of the system, data requirements for training and validation data can be derived (see also \cite{Vogelsang2019}).

\subsection{Threats to validity}
The study focused on context definitions for the development of automated driving systems that use machine learning for the perception system.
Automated driving system are often considered context-aware systems and therefore the findings of this study could potentially be transferred to other context-aware systems. 
Most of the interviews were conducted at one Tier 1 supplier company with offices in Sweden and the United States. 
To support generalisability of the results, a sampling strategy was chosen that included different roles on different levels and at both locations of the case company. 
In addition, individual interviewees outside of the case company were included in the study. 
A threat to validity is the sole focus on the automotive industry. In order to support transferability of the results to other fields, the interview questions were formulated with the intent of being non-specific to the automotive field.

\subsection{Conclusion}
This case study was conducted in the setting of an automotive supplier company by collecting qualitative data through interviews with automotive experts, thematic analysis of the data, and validation of findings through a focus group and background literature triangulation. 
The results show a lack of standardisation of concepts and processes for defining the operational context and deriving the ODD for automated driving systems. Because of the typically distributed development of systems in the automotive industry, this creates challenges which lead to misinterpretation and slow iteration loops between the stakeholders. A major problem the study identifies are missing documentations of context assumptions. Whether a context is assumed, or explicitly given through a use case, can make a difference during the function development and testing. Furthermore, the study reveals a lack of involvement of function developers in the requirement engineering activities that lead to the context definition. As a result, function developers often misinterpret or question the defined operational context.\par
The study also elicited possible solutions to the challenges. Besides obvious solutions, such as more standardisation and deeper involvement of function developers in the definition of the operational context, the participants also suggested ideas such as diverse data about the context, completeness criteria for the ODD, more efficient leveraging from other contexts, and improved integration of context definitions into scaled agile frameworks. These ideas can serve future efforts and research towards a standardisation of context definitions for automated driving systems or other context-aware systems that use AI.
%
%
%
\bibliographystyle{splncs04}
\bibliography{bib}

\begin{thebibliography}{10}
\providecommand{\url}[1]{\texttt{#1}}
\providecommand{\urlprefix}{URL }
\providecommand{\doi}[1]{https://doi.org/#1}

\bibitem{Brown1996}
Brown, P.J.: {The stick-e document: a framework for creating context-aware
  applications}. Tech. Rep.~June, Electronic Publishing-Chichester (1996)

\bibitem{Chen2003}
Chen, H., Finin, T., Joshi, A.: {An ontology for context-aware pervasive
  computing environments}. The Knowledge Engineering Review  \textbf{18}(3),
  197--207 (2003)

\bibitem{Cockburn2000}
Cockburn, A.: {Writing effective use cases}. Tech. rep., Addison-Wesley Longman
  (2000)

\bibitem{Colwell2018}
Colwell, I., Phan, B., Saleem, S., Salay, R., Czarnecki, K.: {An Automated
  Vehicle Safety Concept Based on Runtime Restriction of the Operational Design
  Domain}. Intelligent Vehicles Symposium, Proceedings pp. 1910--1917 (2018)

\bibitem{Creswell2014}
Creswell, J.W.: {Research Design: Qualitative, Quantitative, and Mixed Methods
  Approaches}. Sage Publications, 4th edn. (2014)

\bibitem{Creswell2017}
{Creswell, John W.; Poth}, C.N.: {Qualitative Inquiry and Research Design:
  Choosing Among Five Approaches}. Sage Publishing, fourth edn. (2017)

\bibitem{Czarnecki2018}
Czarnecki, K.: {Operational Design Domain for Automated Driving Systems -
  Taxonomy of Basic Terms} (2018)

\bibitem{Damak2020}
Damak, Y., Leroy, Y., Trehard, G., Jankovic, M.: Operational context-based
  design method of autonomous vehicles logical architectures. In: 15th
  International Conference of System of Systems Engineering (SoSE). pp.
  439--444. IEEE (2020)

\bibitem{Gyllenhammar2020}
Gyllenhammar, M., Johansson, R., Warg, F., Chen, D., Heyn, H.M., Sanfridson,
  M., S{\"{o}}derberg, J., Thors{\'{e}}n, A., Ursing, S.: {Towards an
  Operational Design Domain That Supports the Safety Argumentation of an
  Automated Driving System}. 10th European Congress on Embedded Real Time
  Systems pp. 1--10 (2020)

\bibitem{Henricksen2004}
Henricksen, K., Indulska, J.: {A software engineering framework for
  context-aware pervasive computing}. In: Proceedings of the Second Annual
  Conference on Pervasive Computing and Communications. pp. 77--86. IEEE (2004)

\bibitem{ISO4804}
{International Organization for Standardization}: {ISO/TR 4804:2020 Road
  vehicles — Safety and cybersecurity for automated driving systems —
  Design, verification and validation}. International Organization for
  Standardization, Geneva (2020), \url{www.iso.org}

\bibitem{Irvine2021}
Irvine, P., Zhang, X., Khastgir, S., Schwalb, E., Jennings, P.: A two-level
  abstraction odd definition language: Part i. In: 2021 IEEE International
  Conference on Systems, Man, and Cybernetics (SMC). pp. 2614--2621. IEEE
  (2021)

\bibitem{Jackson1995}
Jackson, M.: The world and the machine. In: 17th International Conference on
  Software Engineering (ICSE). pp. 283--283. IEEE (1995)

\bibitem{Dey2001}
{K. Dey}, A.: {Understanding and using context}. Personal and ubiquitous
  computing pp.~4--7 (2001)

\bibitem{Knauss2016}
Knauss, A., Damian, D., Franch, X., Rook, A., M{\'{u}}ller, H.A., Thomo, A.:
  {Acon: A learning-based approach to deal with uncertainty in contextual
  requirements at runtime}. Information and Software Technology  \textbf{70},
  85--99 (2016)

\bibitem{Knauss2014}
Knauss, A., Damian, D., Schneider, K.: {Eliciting contextual requirements at
  design time: A case study}. In: 4th International Workshop on Empirical
  Requirements Engineering (EmpiRE). pp. 56--63. IEEE (2014)

\bibitem{Koopman2019}
Koopman, P., Fratrik, F.: How many operational design domains, objects, and
  events? In: Proceedings of AAAI Workshop on Artificial Intelligence Safety.
  Honolulu, USA (2019)

\bibitem{Nemoto2015}
Nemoto, Y., Uei, K., Sato, K., Shimomura, Y.: {A Context-based Requirements
  Analysis Method for PSS Design}. Procedia CIRP  \textbf{30},  42--47 (2015)

\bibitem{NHTSA2017}
NHTSA: {Automated Driving Systems: a vision for safety} (2017)

\bibitem{Palinkas2015}
Palinkas, L.A., Horwitz, S.M., Green, C.A., Wisdom, J.P., Duan, N., Hoagwood,
  K.: Purposeful sampling for qualitative data collection and analysis in mixed
  method implementation research. Administration and Policy in Mental Health
  and Mental Health Services Research  \textbf{42}(5),  533--544 (2015).
  \doi{10.1007/s10488-013-0528-y}

\bibitem{Pfeffer2019}
Pfeffer, R., Basedow, G.N., Thiesen, N.R., Spadinger, M., Albers, A., Sax, E.:
  Automated driving - challenges for the automotive industry in product
  development with focus on process models and organizational structure. In:
  2019 International Systems Conference (SysCon). pp.~1--6. IEEE (2019)

\bibitem{Ramirez2012}
Ramirez, A.J., Jensen, A.C., Cheng, B.H.C.: {A taxonomy of uncertainty for
  dynamically adaptive systems}. In: 7th International Symposium on Software
  Engineering for Adaptive and Self-Managing Systems (SEAMS). pp. 99--108. IEEE
  (2012)

\bibitem{Reschka2012}
Reschka, A., B{\"{o}}hmer, J.R., Nothdurft, T., Hecker, P., Lichte, B., Maurer,
  M.: {A surveillance and safety system based on performance criteria and
  functional degradation for an autonomous vehicle}. Conference on Intelligent
  Transportation Systems, Proceedings (ITSC) pp. 237--242 (2012)

\bibitem{SAE2018}
SAE: {J3016B Taxonomy and Definitions for Terms Related to Driving Automation
  Systems for On-Road Motor Vehicles}. Tech. rep., SAE International (2018),
  \url{https://www.sae.org/standards/content/j3016_201806/}

\bibitem{Saldana2013}
Salda{\~{n}}a, J.: {The coding manual for qualitative researchers}. Sage
  Publishing, second edn. (2013)

\bibitem{Schwalb2021}
Schwalb, E., Irvine, P., Zhang, X., Khastgir, S., Jennings, P.: A two-level
  abstraction odd definition language: Part ii. In: 2021 IEEE International
  Conference on Systems, Man, and Cybernetics (SMC). pp. 1669--1676. IEEE
  (2021)

\bibitem{Shalev-Shwartz2017}
Shalev-Shwartz, S., Shammah, S., Shashua, A.: {On a Formal Model of Safe and
  Scalable Self-driving Cars}. arXiv pp. 1--37 (2017)

\bibitem{Soultana2019}
Soultana, A., Benabbou, F., Sael, N.: {Context-awareness in the smart car}. In:
  Proceedings of the 4th International Conference on Smart City Applications
  (SCA). pp.~1--8. ACM Press, New York, New York, USA (2019)

\bibitem{Thorn2018}
Thorn, E., Kimmel, S., Chaka, M.: {A Framework for Automated Driving System
  Testable Cases and Scenarios} (2018),
  \url{https://www.nhtsa.gov/sites/nhtsa.dot.gov/files/documents/13882-automateddrivingsystems_092618_v1a_tag.pdf}

\bibitem{CambridgeContext2021}
{University of Oxford}: {Oxford Learner's Dictionary, Entry: Context} (2021),
  \url{https://www.oxfordlearnersdictionaries.com/definition/english/context}

\bibitem{Vogelsang2019}
Vogelsang, A., Borg, M.: Requirements engineering for machine learning:
  Perspectives from data scientists. In: IEEE 27th International Requirements
  Engineering Conference (RE). pp. 245--251. IEEE (2019)

\end{thebibliography}

\end{document}